# Subluminal to superluminal propagation of an optical pulse in an f-deformed Bose- Einstein condensate

Z. Haghshenasfard[1], M. H. Naderi[2] and M. Soltanolkotabi

Quantum Optics Group, Department of Physics, University of Isfahan, Isfahan, Iran

**Abstract**

In this paper, we investigate the propagation of a weak optical probe pulse in an f-deformed Bose- Einstein condensate (BEC) of a gas with the $\Lambda$-type three- level atoms in the electromagnetically induced transparency (EIT) regime. We use an f- deformed generalization of an effective two- level quantum model of the three- level $\Lambda$ configuration in which the Gardiner's phonon operators for BEC are deformed by an operator- valued function, $f(\hat{n})$, of the particle- number operator $\hat{n}$. With making use of the quantum approach of the angular momentum theory we obtain the eigenvalues and eigenfunctions of the system up to first order approximation. We consider the collisions between the atoms as a special kind of f- deformation. The collision rate $\kappa$ is regarded as the deformation parameter and light propagation in the deformed BEC is analyzed. In particular, we show that the absorptive and dispersive properties of the deformed condensate can be controlled effectively by changing the deformation parameter $\kappa$ and the total number of atoms. We find that by increasing the value of $\kappa$ the group velocity of the probe pulse changes, through deformed condensate, from subluminal to superluminal.

---

[1] E-mail: zhaghshenas@hotmail.com
[2] E-mail: mhnaderi2001@yahoo.com

# 1 Introduction

In the past few decades, there have attracted tremendous interests in the development of theoretical studies and experimental techniques for controlling the light pulse propagation through resonant optical media. Manipulation of light propagation in an optical medium can be realized by changing the dispersive properties of the medium. The exact control over the optical properties of the medium such as dispersion, absorption and refraction index gives rise to observation of some fascinating phenomena such as coherent population trapping (CPT) [1], lasing without inversion (LWI) [2], electromagnetically induced transparency (EIT) [3, 4], subluminal and superluminal propagation [5, 6, 7].

In recent years, there have been many studies of both subluminal and superluminal light propagation in atomic media having very special controlled optical properties. In general, subluminal propagation takes place at the normal dispersive regions where the group velocity is smaller than the vacuum speed of light $c$. Slow group velocity in coherent media has been shown to provide new regimes of nonlinear interacting with highly increased efficiency even for very weak light fields. The underlying mechanism of the subluminal propagation is EIT [8, 9], in which the coupling laser field alters the optical properties of the medium. It has been demonstrated [4, 10] that EIT is accompanied by large frequency dispersion, thus the group velocity can be slow down up to $10 - 10^2 \, m/s$ [7, 11, 12]. Harris and co-workers have found [11] that large positive dispersion of refractive index in the EIT window can be used to reduce dramatically the group velocity of light pulses which leads to nonlinear optics at low light level [13] and quantum memory [14]. Furthermore, Kocharovskaya et al [15] have shown how the spatial dispersion can even stop light in a hot gas. Using EIT and adiabatic following of dark state polaritons, the group velocity of light pulses can be dramatically decelerated and their quantum state can be mapped into a metastable collective state of atomic ensembles [16] which is of application in quantum information. Slow light has also found many potential applications in optical delay lines [7], quantum entanglement of slow photons [9], non classical (e. g. squeezed) and entangled atomic ensembles [13] and optical black hole [17]. On the other hand, superluminal propagation occurs in optical media with anomalous dispersion [6, 18]. Superluminal group velocity has restarted the debate about the definitions of information velocity. On this issue, it has been clarified [19] that the information velocity, i. e., the speed of a point of non-analytically, can never exceed in vacuum speed of light, not even in the case of superluminal group velocity and causality is preserved. Wang et al. [20] demonstrated superluminal light propagation using the region of lossless anomalous dispersion between two closely spaced gain lines in a double-peaked Raman gain medium. The gain doublet is created by applying two intense detuned cw pumps with slightly different frequencies to one transition of a $\Lambda$-type three- level system in atomic cesium. Akulshin and his coworkers demonstrated [21] that extraordinary steep anomalous dispersion can be obtained in two-level atomic systems by using electromagnetically induced absorption (EIA). They observed a value of anomalous dispersion corresponding to a negative group velocity $v_g \cong -\dfrac{c}{23000}$ with large absorption. Steep anomalous dispersion in the absence of absorption has been shown both theoretically and experimentally in a two- level atomic system strongly driven by a resonant pump [22, 23], for both weak and moderately strong probes [23]. The superluminal propagation of pulses has also been studied inside diffractive structures [24], photonic band gap materials [25], active plasma medium [26] and nonlinear coherent medium [27]. The possible applications of superluminal light propagation can be found in optical communication, optical networks, opto- electronic devices and wave guides. In particular, by using the principle of superluminality one can speed up the clock of a computer chip [28].

In view of the many potential applications of subluminal and superluminal light propagation, there have been considerable, both theoretical and experimental, attempts to realize both superluminal and subluminal light in a single system. Recently light propagation from subluminal to superluminal in $\Lambda$- type and V- type three- level atoms has been shown [29, 30,31, 32, 33]. Agarwal et al [30] proposed the idea of obtaining the light propagation changing from subluminal to superluminal group velocity by applying a coupling filed connecting the two lower metastable states of a $\Lambda$ system. Sahrai et al [31] have shown tunable phase control from subluminal to superluminal light propagation in an $\Lambda$- type atom with an extra energy level. Furthermore, Carreno et al [32] have shown theoretically the possibility of light propagation form subluminal to superluminal in $\Lambda$- type atoms with closely spaced lower levels, when the atom is driven by a strong coherent field and damped by a broad band squeezed vacuum. In a V- type system, a method for manipulating the group velocity of weak pulse from subluminal to superluminal by adjusting the relative phase of the probe and pump fields has also been proposed [33]. Experimental evidences of fast and slow light propagation in four- level atoms have also been obtained [34].

One of the most important problems in EIT was the spectrum broadening due to thermal effects. To avoid this problem a highly coherent atomic Bose- Einstein condensate (BEC) is used [7, 35]. BEC is a purely quantum mechanical phenomenon. It is a phase transition occurring at sufficiently low temperature for a gas composed by bosons, when the thermal de Broglie wavelength grows to be of the same order as the interparticle distance. Under such conditions the spectrum broadening regarded to thermal effects become less significant. Then EIT in a BEC is an excellent environment for studying the subluminal pulse. By using this technique Hau and coworkers [7] demonstrated the ultra slow group velocities of $17 \, m/s$ in an ultracold gas of sodium atoms. Based on their results we consider the EIT effects in BEC of $\Lambda$- type configuration sodium atoms in the limit of particle- number conservation. To this end, we use a description in terms of deformed bosons. As is known, to study the dynamics of BEC gas, the Bogolubov approximation in quantum many body theory is usually applied, in which the creation and annihilation operators for condensate atoms are replaced by a c- number. But, the Bogolubov approximation destroys the conservation of the total particle number. To overcome this problem, Gardiner [36] suggested a modified Bogolubov approximation by introducing phonon operators which conserve the total atomic particle number $N$ and obey the f- deformed commutation relation of Heisenberg- Weyl algebra such that as $N \to \infty$, the usual commutation relation of Heisenberg- Weyl algebra is regained. Then the Gardiner's phonon approach gives an elegant infinite atomic particle- number approximation theory for BEC taking into account the conservation of the total atomic number [37].

The concept of the f- deformed boson has been extensively applied in several physical models. One of the most important properties of f- deformed bosons is their relation to nonlinearity of a special type and it is shown that [38] some nonlinear dynamical systems are related to deformation of linear classical and quantum systems. For example, the q- deformed oscillator has been interpreted as a nonlinear oscillator with a special type of non-linearity which classically corresponds to an intensity dependence of the oscillator frequency [38]. The relation between deformed radiation field and nonlinear quantum optical processes has also been studied in [39]. If the f- nonlinearity is considered, many objects may be defined by using the concept of the f- deformed bosons [40]. For example, f- nonlinearity deforms the formula for the mean photon number in black body relation [38], this nonlinearity changes the specific heat behavior and for small temperature, the behavior of the deformed Planck distribution differs form the usual one. Another possible phenomena related to the f- nonlinearity has been considered in [38]. Recently, much attention has been paid to understanding and applications of the f- deformed boson for the description of BEC [41, 42].

In this paper, we consider a deformed condensate in which we can control the propagation of light from subluminal to superluminal by changing the deformation parameters. The system under consideration is an f- deformed BEC of a gas with $\Lambda$- type three- level atoms in the EIT regime, in which the Gardiner's phonon operators for BEC are deformed by an operator- valued function $f(\hat{n})$. By considering the effect of collision between the atoms within the condensate as a special kind of the f- deformation, in which the collision rate $\kappa$, is regarded as deformation parameter, we analyze the light propagation in the deformed BEC. Such a system offers extra degrees of flexibility $(\kappa, N)$ for processing signal and an effective control of the group velocity of light in deformed BEC can be demonstrated by changing the controlling parameters $(\kappa, N)$ which leads to both subluminal and superluminal propagation of the light. We show that the f- deformed BEC exhibits nonlinear characteristics, such that the nonlinearity increases by adjusting deformation parameters $(\kappa, N)$, and the nonlinearity improve the subluminal and superluminal properties of the light.

The paper is organized as follows. In section 2 we present our model and we use the deformed algebra to study the condensate with large but finite number of atoms. We show that a physical and natural realization of the f- deformed boson is provided by the Gardiner's phonon operator, for the description of the BEC. Here the deformation parameter is no longer phenomenological and is defined by the total atom number. We show that the effect of collisions between the atoms within condensate is an extra deformation on the intrinsically deformed Gardiner's phonon operators for BEC. In section 3 the quantum approach of the angular momentum is used to obtain the eigenvalues and eigenfunctions of the system up to first- order approximation. The interaction between the f- deformed BEC and the probe pulse and conditions for subluminality and superluminality of the probe pulse are studied in section 4. Finally we summarize our results in section5.

## 2 The effective two- level model and f- deformed bosonic algebra for the Gardiner's phonon

We consider the EIT effects in the BEC of $\Lambda$- type three- level atoms with energy level $E_2 < E_1 < E_3$, interacting with two laser fields (Fig. 1). The lower two levels $|1\rangle$ and $|2\rangle$ are coupled to the upper level $|3\rangle$. We call the field with frequency $\omega_p$, the probe field and the field with $\omega_c$, the coupling field. Under rotating wave approximation the total Hamiltonian of the system is given by

$$\hat{H} = \hbar\omega_{12}|1\rangle\langle 1| + \hbar\omega|3\rangle\langle 3| + \hbar[(-g_1|3\rangle\langle 1|e^{-i\omega_c(t-\frac{z}{c})} - g_2|3\rangle\langle 2|e^{-i\omega_p(t-\frac{z}{c})}) + H.C], \quad (1)$$

where coupling constants are defined by $g_1 = \dfrac{\mu_{31}A_c}{\hbar}$ and $g_2 = \dfrac{\mu_{32}A_p}{\hbar}$ with $\mu_{ij}$ denoting the transition dipole matrix element between states $|i\rangle$ and $|j\rangle$ and $A_{c(p)}$ being slow varying coupling (probe) field amplitudes.

We denote single atom density operator by $\hat{\rho} = \sum_{i,j=1}^{3} \rho_{ij}|i\rangle\langle j|$. The time evolution of $\hat{\rho}$ is described by the Liouville equation of motion [43]

$$\frac{\partial\hat{\rho}}{\partial t} = -\frac{i}{\hbar}[\hat{H},\hat{\rho}] - \gamma_{12}(|1\rangle\langle 1|\hat{\rho} - 2|2\rangle\langle 1|\hat{\rho}|1\rangle\langle 2| + \hat{\rho}|1\rangle\langle 1|$$
$$-\gamma_{32}(|3\rangle\langle 3|\hat{\rho} - 2|2\rangle\langle 3|\hat{\rho}|3\rangle\langle 2| + \hat{\rho}|3\rangle\langle 3| - \gamma_{31}(|3\rangle\langle 3|\hat{\rho} - 2|1\rangle\langle 3|\hat{\rho}|3\rangle\langle 1| + \hat{\rho}|3\rangle\langle 3|. \quad (2)$$

Here, the constants $\gamma_{ij}$ determine the rate of spontaneous decay from the level $|i\rangle$ to level $|j\rangle$ in the $\Lambda$- scheme. Since we want to study nonlinear interaction of the BEC with the probe pulse, we consider only the explicit dependence of $\bar{\rho}_{32}$ on the Rabi frequency $g_2$ of the probe pulse [43]:

$$\bar{\rho}_{32} = \bar{\rho}_{32}^{(0)} + \bar{\rho}_{32}^{(1)}g_2 + \bar{\rho}_{32}^{(2)}|g_2|^2 + \bar{\rho}_{32}^{(3)}|g_2|^2 g_2, \quad (3)$$

where $\bar{\rho}$ denotes the averaged density operator over the rapidly oscillating phase of the fields. In Eq. (3) $\bar{\rho}_{32}^{(0)}$ denotes initial polarization in the BEC, $\bar{\rho}_{32}^{(1)}$ corresponds to the stationary solution of the system for the linear susceptibility of the BEC and the nonlinear corrections $\bar{\rho}_{32}^{(2)}$ and $\bar{\rho}_{32}^{(3)}$ determine resonant nonlinear atomic susceptibility. The calculation shows that $\bar{\rho}_{32}^{(2)}$ is negligible comparing with the Kerr type nonlinearity $\bar{\rho}_{32}^{(3)}$, thus we neglect this term. We consider the dynamics of the levels $|2\rangle$ and $|3\rangle$ of three- level $\Lambda$- type atoms interacting with the probe field in the EIT regime. Assuming all atoms being initially in the ground state $|2\rangle$ one can find [43]

$$\bar{\rho}_{32}^{(1)} = \frac{1}{\Gamma}, \bar{\rho}_{32}^{(3)} = \frac{i}{\Gamma}\frac{\Gamma^* - \Gamma}{2|\Gamma^2|}(\frac{1}{2\gamma_{opt}} + \frac{1}{\gamma_{mag}}), \quad (4)$$

where

$$\Gamma = \Delta - 2i\gamma_{opt} + \frac{|g_1|^2}{i\gamma_{mag} - \Delta},$$

$$\gamma_{opt} = \frac{\gamma_{31} + \gamma_{32}}{2}, \gamma_{mag} = \gamma_{12}.$$

(5)

We introduce the creation and annihilation operators $\hat{a}^+$ and $\hat{a}$, respectively, for the probe field and assuming strong coupling field so that its intensity is given by a c- number parameter. Under the rotating wave and dipole approximations the Hamiltonian (1) can be written as an effective two- level Hamiltonian in the following form [43]

$$\hat{H} = \hbar[\omega_p(\hat{a}^+\hat{a} + \hat{S}_3 + \frac{\hat{N}}{2}) + \Delta\hat{S}_3 + k_1(\hat{a}\hat{S}_+ + \hat{a}^+\hat{S}_-) + k_2(\hat{a}^+\hat{a}\hat{a}^+\hat{S}_- + \hat{a}\hat{a}^+\hat{a}\hat{S}_+)], \quad (6)$$

where $\hat{N}$ is the total number of atoms in BEC and the operators $S_\pm, S_3$ describe total dipole momentum corresponding to the transitions $|3\rangle \to |2\rangle$ for the atoms in BEC. The first and the second terms in Eq. (6) give the free energy of the probe field and atoms, respectively. The third and the last terms describe linear and nonlinear contributions into the interaction between the probe field and two-level particles, respectively. The two coupling constants are defined by [43]:

$$k_1 = k_0 L_l, \quad k_2 = k_0^3 L_{nl}, \quad (7)$$

where, $k_0 = \mu_{32}\sqrt{\frac{\omega_p}{2\hbar\varepsilon_0 V}}$, is the single-photon Rabi-frequency in the Dicke model [44], and the parameters $L_{l(nl)}$ denote linear and nonlinear coupling constants, respectively.

$$L_l = \frac{\rho_{32}^{(1)}(g_1,\Delta)}{\rho_{32}^{(1)}(g_1=0,\Delta)}, L_{nl} = \frac{\rho_{32}^{(3)}(g_1,\Delta)}{\rho_{32}^{(1)}(g_1=0,\Delta)}. \quad (8)$$

We now apply the second quantization theory to the above model. Let $\hat{b}_{e(g)}, \hat{b}_{e(g)}^+$ denote the annihilation and creation operators of the atoms in the excited (ground) state, respectively, so that the total Hamiltonian reads as

$$\hat{H} = \hbar[\omega_p\{\hat{a}^+\hat{a} + (\hat{b}_e^+\hat{b}_e - \hat{b}_g^+\hat{b}_g) + \frac{(\hat{b}_e^+\hat{b}_e + \hat{b}_g^+\hat{b}_g)}{2}\} + \Delta(\hat{b}_e^+\hat{b}_e - \hat{b}_g^+\hat{b}_g)$$
$$+ k_1(\hat{a}\hat{b}_e^+\hat{b}_g + \hat{a}^+\hat{b}_g^+\hat{b}_e) + k_2(\hat{a}^+\hat{a}\hat{a}^+\hat{b}_g^+\hat{b}_e + \hat{a}\hat{a}^+\hat{a}\hat{b}_e^+\hat{b}_g)]. \quad (9)$$

It should be noted that in the above Hamiltonian the total atom number $\hat{N} = \hat{b}_e^+\hat{b}_e + \hat{b}_g^+\hat{b}_g$ is conserved $([\hat{N}, \hat{H}] = 0)$. In the thermodynamic limit $(\hat{N} \to \infty)$, the Bogolubov approximation is usually applied, in which the ladder operators $\hat{b}_g^+, \hat{b}_g$ of the ground state are replaced by a c-number $\sqrt{N_c}$, where $N_c$ is the average number of the initial condensated atoms. In this limit, Hamiltonian (9) describes a system of two coupled harmonic oscillators

$$\hat{H}_b = \hbar\omega_p \hat{a}^+\hat{a} + \hbar[\frac{3\omega_p}{2} + \Delta]\hat{b}_e^+\hat{b}_e +$$
$$\hbar k_1(a\sqrt{N_c}\hat{b}_e^+ + a^+\sqrt{N_c}\hat{b}_e) + \hbar k_2(\hat{a}^+\hat{a}\hat{a}^+\sqrt{N_c}\hat{b}_e + \hat{a}\hat{a}^+\hat{a}\sqrt{N_c}\hat{b}_e^+).$$
(10)

The third term in Eq. (10) describes the interaction between the field and two-level particles and it has the typical form of the Dicke model [44]. The last term in Eq. (10) describes nonlinear processes due to the presence of strong classical coupling field, and it depends on the intensity of the probe pulse. However, this approximation destroys symmetry of the Hamiltonian (9), i.e., the conversation of the total particle number is violated ($[\hat{N}, \hat{H}_b] \neq 0$). In order to preserve the property of the initial model, it is possible to introduce the following Gardiner's phonon operators [36].

$$\hat{b}_q = \frac{1}{\sqrt{N}}\hat{b}_g^+\hat{b}_e, \quad \hat{b}_q^+ = \frac{1}{\sqrt{N}}\hat{b}_g\hat{b}_e^+. \tag{11}$$

These operators obey a deformed algebra. In fact, a straightforward calculation leads to the following commutation relation

$$[\hat{b}_q, \hat{b}_q^+] = 1 - \frac{2}{N}\hat{b}_e^+\hat{b}_e = 1 - 2\eta\hat{b}_e^+\hat{b}_e, \tag{12}$$

where the operator parameter $\eta = \frac{1}{N}$ is introduced. For sufficiently large number of atoms the $\eta$ parameter is considered as a c-number. The algebra defined by Eq.(12) belongs to the f-deformed algebra [37], where in general the deformed operator is related to nondeformed one through an operator valued function $f_1$ as

$$\hat{b}_q = \hat{b}f_1(\hat{b}_e^+\hat{b}_e; \eta). \tag{13}$$

In our particular case, we have

$$f_1(\hat{b}_e^+\hat{b}_e; \eta) = \sqrt{1 - \eta(\hat{b}_e^+\hat{b}_e - 1)}. \tag{14}$$

It is obvious that in the limiting case $\eta \to 0 (N \to \infty)$ the deformation function tends to unity. It means that the influence of deformation reduces with increasing $N$. The above argument reveals that by introducing the Gardiner's phonon operators we arrive at an intrinsically deformed model in which the deformation parameter is determined by the total number $N$. For small deformation, the deformed boson operators $\hat{b}_q^+$ and $\hat{b}_q$ could be expressed in terms of the normal bosonic operators $\hat{b}^+$ and $\hat{b}$ ($[\hat{b}, \hat{b}^+] = 1$) as

$$\hat{b}_q \approx \hat{b}[1 - \frac{\eta}{2}(\hat{b}^+\hat{b} - 1)] = \hat{b} - \frac{1}{2N}\hat{b}^+\hat{b}\hat{b},$$
$$\hat{b}_q^+ \approx [1 - \frac{\eta}{2}(\hat{b}^+\hat{b} - 1)]\hat{b}^+ = \hat{b}^+ - \frac{1}{2N}\hat{b}^+\hat{b}^+\hat{b}. \tag{15}$$

Having this in mind, the total Hamiltonian (9) can be written as

$$\hat{H} = \hbar\omega_p \hat{a}^+ \hat{a} + \hbar[\frac{3\omega_p}{2} + \Delta]\hat{b}_q^+ \hat{b}_q + \quad (16)$$

$$\hbar K_1(\hat{a}\hat{b}_q^+ + \hat{a}^+\hat{b}_q) + \hbar K_2(\hat{a}^+\hat{a}\hat{a}^+\hat{b}_q + \hat{a}\hat{a}^+\hat{a}\hat{b}_q^+),$$

where $K_1 = \sqrt{N}k_1, K_2 = \sqrt{N}k_2.$ (17)

Now, we propose to illustrate the effect of collisions between the atoms within condensate as a special kind of f- deformation. For this, we briefly review the deformed oscillator algebra in the following.

## 2-1 Deformed oscillator algebra

As we know, the q- deformed oscillator algebra is a three- element Lie algebra $\{\hat{A}, \hat{A}^+, \hat{N}\}$ plus one- parameter $q$, which modifies the commutation relations. Even for single parameter, there exist many different modifications. One of the most frequently examples is a four- parameter generalized deformed algebra which was introduced in [45] on the basis of the definition of the deformation

$$\hat{A}\hat{A}^+ + q\hat{A}^+\hat{A} = q^{\alpha\hat{N}+\beta} \qquad [\hat{N},\hat{A}] = -\hat{A} \qquad [\hat{N},\hat{A}^+] = -\hat{A}^+. \quad (18)$$

Man'ko et al [38] introduced a realization for the operators $\hat{A}$ and $\hat{A}^+$ in terms of the so- called f- oscillators, defined as a nonlinear expansion of the usual harmonic oscillator operators $\hat{a}$ and $\hat{a}^+$,

$$\hat{A} = \hat{a}f(\hat{N}), \hat{A}^+ = f^*(\hat{N})\hat{a}^+, \hat{N} \equiv \hat{a}^+\hat{a}. \quad (19)$$

The function $f(\hat{N})$ is specific to each q- deformed algebra, and depends on four parameters $\alpha, \beta, \gamma$ and $q$. The Heisenberg equation of motion for $\hat{A}$ is

$$\dot{\hat{A}} + i\omega(\hat{N})\hat{A} = 0, \quad (20)$$

where the nonlinear frequency $\omega(\hat{N})$ is defined from

$$[\hat{A}, \hat{H}(\hat{N})] = \omega(\hat{N})\hat{A}. \quad (21)$$

According to the commutation relation (18), $\hat{H}(\hat{N})$ and $\omega(\hat{N})$ are related by

$$\hat{H}(\hat{N}+1) - \hat{H}(\hat{N}) = \omega(\hat{N}). \quad (22)$$

As an example, let us consider the Hamiltonian for the free f- oscillator

$$\hat{H}(\hat{N}) = \frac{\hbar\omega_0}{2}(\hat{A}^+\hat{A} + \hat{A}\hat{A}^+) = \frac{\hbar\omega_0}{2}[\left|f(\hat{N}+1)\right|^2(\hat{N}+1) + \left|f(\hat{N})\right|^2\hat{N}], \quad (23)$$

where $\hat{N}$ is a constant of motion ($[\hat{N},\hat{H}] = 0$) and the frequency can also express as

$$\omega(\hat{N}) = \frac{\hbar\omega_0}{2}[\left|f(\hat{N}+2)\right|^2(\hat{N}+2) - \left|f(\hat{N})\right|^2\hat{N}]. \quad (24)$$

Using deformation (19) and the algebra (18) we find

$$\left|f(\hat{N})\right|^2 = \begin{cases} \dfrac{q^\beta}{\hat{N}}\dfrac{q^\alpha - q^{\gamma\hat{N}}}{q^\alpha - q^\gamma} & \alpha \neq \gamma \\ q^{\beta+\gamma(\hat{N}-1)} & \alpha = \gamma \end{cases}. \quad (25)$$

By introducing new deformation parameters using relations $q = e^{\tau}, \alpha = \upsilon + \mu, \gamma = \upsilon - \mu$, Eq. (25) becomes

$$\left|f(\hat{N})\right|^2 = \frac{\sinh(\tau\mu\hat{N})}{\hat{N}\sinh(\tau\mu)}\exp\{\tau[\beta + \upsilon(\hat{N}-1)]\}, \tag{26}$$

and the Hamiltonian of the free f- oscillator (23) can be written explicitly as

$$\hat{H} = \frac{\hbar\omega_0}{2}e^{\tau(\beta+\upsilon\hat{N})}\{\frac{\sinh(\tau\mu[\hat{N}+1])}{\sinh(\tau\mu)} + e^{-\tau\upsilon}\frac{\sinh(\tau\mu\hat{N})}{\sinh(\tau\mu)}\}. \tag{27}$$

For the special case $\mu = 0$ (i. e. $\alpha = \gamma$) and $\beta = 0$, we obtain the following expression for the eigenvalues $E_n(\hat{H}|n\rangle = E_n|n\rangle)$,

$$E_n \approx \frac{\hbar\omega_0}{2}e^{\tau\upsilon n}[1 + n(1 + e^{-\tau\upsilon})]. \tag{28}$$

And the frequency is

$$\omega(n) \approx \hbar\omega_0 e^{\tau\upsilon n}[e^{\tau\upsilon} + n\sinh(\tau\upsilon)]. \tag{29}$$

We can gain a better insight into the effects of deformation by assuming $\tau = 1$ and $\upsilon n \ll 1$, and keeping terms up to $n^2$, in such case we obtain

$$E_n = \frac{\hbar\omega_0}{2}[1 + n(1 + \upsilon + e^{-\upsilon}) + n^2\upsilon(1 + e^{-\upsilon})], \tag{30}$$

which is characteristic of a collisions effect between the atoms within BEC, and

$$\omega_n \approx \omega_0[e^{\upsilon} + n\upsilon(1 + e^{\upsilon}) + n^2\upsilon^3]. \tag{31}$$

So, even for small nonlinearity energy and frequency are not proportional.

**2-2 An example: the collisions effect**

As a particular physical example, we consider the effect of collisions between the atoms within the condensate. The effective interaction Hamiltonian contains a nonlinear term proportional to $(\hat{a}^+\hat{a})^2$ [46],

$$\hat{H}_I = \frac{\hbar\kappa}{2}(\hat{a}^+\hat{a})^2, \tag{32}$$

where the collision rate is denoted by $\kappa$. By expanding the Hamiltonian (23) and considering small values of $\upsilon$ and $\mu^2$ we obtain

$$\hat{H}(\hat{N}) = \frac{\hbar\omega_0}{2}[(2\hat{N}+1) + \frac{1}{6}\mu^2\hat{N} + (\frac{1}{2}\mu^2 + 2\upsilon)\hat{N}^2 + O(\upsilon^2, \upsilon\mu^2, \mu^4)]. \tag{33}$$

where the interaction Hamiltonian reads as

$$\hat{H}_I(\hat{N}) \approx \frac{\hbar\omega_0}{2}[(\frac{1}{6}\mu^2\hat{N} + (\frac{1}{2}\mu^2 + 2\upsilon)\hat{N}^2]. \tag{34}$$

Since the nonlinear term in Hamiltonian (34) contains only one parameter, the Hamiltonian (34) reproduces the Hamiltonian (32) by setting $\mu^2 = 0$ and $\upsilon = \frac{\kappa}{2\omega_0}$. Thus, we see that the collision effect transforms the standard (nonlinear) harmonic oscillator model into an f- deformed one. Alternatively, we could set, *ab initio* in Eq. (23), $f(\hat{N}) = \sqrt{\kappa \hat{N} + (1-\kappa)}$ to obtain (32). Therefore, the parameters of the generalized deformed algebra are related to the rate of atomic collisions $\kappa$.

Subsequently, by considering the effects of collisions between the atoms within condensate, we can apply the extra deformation on the intrinsically deformed Gardiner's phonon operators for BEC by an operator- valued function $f_2(\hat{n}) = \sqrt{\kappa \hat{n} + (1-\kappa)}$ of the particle number operator $\hat{n}$. Here the nonlinearity is related to the collisions between the atoms within condensate. The operator valued- function $f_2(\hat{n})$ reduces to 1 as soon as $\kappa \to 0$. It means that the deformation increases with the collision rate $\kappa$. In sensible experimental situation, for a condensate composed of rubidium- 87 atoms at temperature of 180 nK, with a density $\rho = 10^{12} \, cm^{-3}$ [41] we can estimate the collision rate to be $\kappa \approx \rho \pi a^2 v_{rms}$, where $a$ is the scattered length and $v_{rms}$ is the root- mean square- speed of the rubidium atoms. We obtain a collision rate of about one collision per second for these parameters, while we can adjust the value of the collision rate $\kappa$ by changing temperature. For example, by decreasing the temperature of the condensate, the velocity $v_{rms}$ of atoms reduces which leads to reduction of collision rate $\kappa$.

The deformed Gardiner's phonon operators are related to the nondeformed ones thorough the operator valued- function $f_2(\hat{n})$ as

$$\hat{B}_q = \hat{b}_q f_2(\hat{n}), B_q^+ = f_2^+(\hat{n})\hat{b}_q^+,$$
$$\hat{n} = \hat{N}_q = \hat{b}_q^+ \hat{b}_q. \tag{35}$$

For small deformation, the f- deformed Gardiner's phonon operators $B_q, B_q^+$ can be expressed as

$$\hat{B}_q = \hat{b}_q[1 - \frac{\kappa}{2}(1 - \hat{b}_q^+ \hat{b}_q)], \quad \hat{B}_q^+ = [1 - \frac{\kappa}{2}(1 - \hat{b}_q^+ \hat{b}_q)]\hat{b}_q^+. \tag{36}$$

Here, the small value for $\kappa$ is considered, where $\kappa(1-\hat{n})$ would be very smaller than one. In that case, by keeping only the lowest order of $\eta = \frac{1}{N}$ for very large total atom number $N$ and by keeping only the first- order term of the collision rate $\kappa$ for very low temperature we get

$$\hat{B}_q \approx (\hat{b} - \frac{1}{2N}\hat{b}^+\hat{b}\hat{b})[1 - \frac{\kappa}{2}\{1 - (\hat{b}^+\hat{b} + \frac{1}{N}\hat{b}^+\hat{b}^+\hat{b}\hat{b})\}],$$
$$\hat{B}_q^+ \approx [1 - \frac{\kappa}{2}\{1 - (\hat{b}^+\hat{b} + \frac{1}{N}\hat{b}^+\hat{b}^+\hat{b}\hat{b})\}](\hat{b}^+ - \frac{1}{2N}\hat{b}^+\hat{b}^+\hat{b}). \tag{37}$$

Therefore, by using (37) the deformed version of the Hamiltonian (16) can be expressed in terms of the nondeformed operators $\hat{b}$ and $\hat{b}^+$ as follows

$$\hat{H} = \hat{H}_0 + \hat{H}', \tag{38}$$
where

$$\hat{H}_0 = \hbar\omega_p(\hat{a}^+\hat{a} + \hat{b}^+\hat{b}) + \hbar K_1(\hat{a}\hat{b}^+ + \hat{a}^+\hat{b}), \tag{39}$$

and

$$\hat{H}' = \hbar[\frac{\omega_p}{2} + \Delta](\hat{b}^+\hat{b}) + \hbar[\frac{3\omega_p}{2} + \Delta](-\frac{1}{N} + \kappa)\hat{b}^+\hat{b}^+\hat{b}\hat{b}$$

$$+ \hbar K_1(\frac{\kappa}{2} - \frac{1}{2N})(a\hat{b}^+\hat{b}^+\hat{b} + \hat{a}^+\hat{b}^+\hat{b}\hat{b}) + \tag{40}$$

$$+ \hbar K_2[(\hat{a}\hat{a}^+\hat{a}\hat{b}^+ + \hat{a}^+\hat{a}\hat{a}^+\hat{b}) + (\frac{\kappa}{2} - \frac{1}{2N})(\hat{a}\hat{a}^+\hat{a}\hat{b}^+\hat{b}^+\hat{b} + \hat{a}^+\hat{a}\hat{a}^+\hat{b}^+\hat{b}\hat{b})].$$

It is evident that the second term in Eq. (40) describes the attractive exciton- exciton collisions due to bi- exciton effect and the third and last terms of $\hat{H}'$ describe the decrease of the exciton– photon coupling constants due to the phase- space filling effect [47]. The forth term of $\hat{H}'$ describes nonlinear processes due to the presence of strong classical coupling field and it depends on the intensity of the probe pulse. It is evident that the Hamiltonian (40) reproduces the Hamiltonian (10) by setting both $\kappa$ and $\eta$ equal to zero. In this case we have neither bi- exciton effect nor phase- space filling effect.

## 3 Approximate analytical solutions

To solve the Schrödinger equation governed by the Hamiltonian (38) we shall make use of the quantum angular momentum theory. According to the Schwinger representation of the angular momentum, we can build the angular momentum operators as follows

$$\hat{J}_z = \frac{1}{2}(\hat{a}^+\hat{a} - \hat{b}^+\hat{b}),$$

$$\hat{J}_+ = \hat{a}^+\hat{b}, \tag{41}$$

$$\hat{J}_- = \hat{a}\hat{b}^+.$$

By using the cavity- field ladder operators $\hat{a}$ and $\hat{a}^+$ and the exciton operators $\hat{b}$ and $\hat{b}^+$ we define

$$\hat{J}_x = \frac{1}{2}(\hat{a}^+\hat{b} + \hat{a}\hat{b}^+), \hat{J}_y = \frac{1}{2i}(\hat{a}^+\hat{b} - \hat{a}\hat{b}^+). \tag{42}$$

Then we rewrite the Hamiltonian (39) as

$$\hat{H}_0 = \hbar\omega_p \hat{N} + 2\hbar K_1 \hat{J}_x = \hbar\omega_p \hat{N} + 2\hbar K_1 e^{-i(\pi/2)\hat{J}_y} \hat{J}_z K_1 e^{+i(\pi/2)\hat{J}_y}, \tag{43}$$

The excitation number operator $N = \hat{b}^+\hat{b} + \hat{a}^+\hat{a}$ is a constant under any SO(3) rotation and $\hat{J}^2 = \hat{J}_x^2 + \hat{J}_y^2 + \hat{J}_z^2 = \frac{\hat{N}}{2}(\frac{\hat{N}}{2} + 1)$ is the total angular momentum operator. The common eigenstates of $\hat{J}^2$ and $\hat{J}_z$ are as follows

$$|jm\rangle = \frac{(\hat{b}^+)^{j-m}(\hat{a}^+)^{j+m}}{\sqrt{(j+m)!(j-m)!}}|0\rangle, \tag{44-a}$$

or

$$|n, n_e\rangle = \frac{(\hat{b}^+)^{n_e} (\hat{a}^+)^n}{\sqrt{n_e! n!}} |0\rangle, \tag{44-b}$$

where $n$ denotes the number of photons, $n_e$ is the number of excitons, and the eigenvalues of $J^2$ and $J_z$ are, respectively

$$j = -\frac{N}{2}, m = -\frac{N}{2}, ..., \frac{N}{2}. \tag{45}$$

The eigenvectors $|\psi_{jm}^0\rangle$ and the eigenvalues $E_{jm}^{(0)}$ of $H_0$ can be easily constructed as

$$|\psi_{jm}^0\rangle = e^{-i(\pi/2)\hat{J}_y} |jm\rangle, \qquad E_{jm}^{(0)} = \hbar\omega_p N + 2\hbar K_1 m. \tag{46}$$

Up to the first-order approximation, the eigenvalues of the Hamiltonian $\hat{H}$ are obtained as

$$E_{jm} = E^{(0)}{}_{jm} + \langle jm| e^{i(\pi/2)\hat{J}_y} \hat{H}' e^{-i(\pi/2)\hat{J}_y} |jm\rangle, \tag{47}$$

and their corresponding eigenfunctions are given by

$$|\psi_{jk}\rangle = |\psi_{jk}^{(0)}\rangle + \sum \frac{\langle jn|\hat{H}'|jk\rangle}{E_{jk}^{(0)} - E_{jn}^{(0)}} |\psi_{jn}^{(0)}\rangle. \tag{48}$$

The matrix elements of the perturbation Hamiltonian $\hat{H}'$ are given by

$$\begin{aligned}
\langle jm| e^{i(\pi/2)\hat{J}_y} \hat{H}' e^{-i(\pi/2)\hat{J}_y} |jm\rangle &= \hbar(\frac{\omega_p}{2} + \Delta)[\frac{1}{2}(j+m) + \frac{1}{2}(j-m)] + \\
&\hbar(\frac{3\omega_p}{2} + \Delta)(-\frac{1}{N} + \kappa)[\frac{1}{4}(j-m)(j-m-1) + \frac{1}{4}(j+m)(j+m-1) + (j^2 - m^2)] + \\
&\hbar K_1 (\frac{\kappa}{2} - \frac{1}{2N})[(\frac{1}{2}(j-m)(j-m-1) + \frac{1}{2}(j+m)(j+m-1)] + \\
&\hbar K_2 [\frac{1}{2}(j+m) - \frac{1}{2}(j-m) + \frac{1}{2}(j+m)^2 - \frac{1}{2}(j-m)^2 + \\
&(\frac{\kappa}{2} - \frac{1}{2N})\{\frac{1}{2}(j+m)(j+m-1) - \frac{1}{2}(j+m)(j+m-1) + \\
&\frac{1}{4}(j+m)(j+m-1)(j+m-2) - \frac{1}{4}(j-m)(j-m-1)(j-m-2) \\
&-\frac{1}{4}(j+m)(j-m)(j+m-1) + \frac{1}{4}(j+m)(j-m)(j-m-1)\}].
\end{aligned} \tag{49}$$

We shall apply this result in section 4 to drive the polarization of the medium.

## 4 Pulse light propagation in the f- deformed BEC

Usually, the interaction of electromagnetic fields with an atomic medium is described by means of the Maxwell- Bloch equations for the atomic density matrix elements, which are solved in the steady state conditions. However, when the number of interacting fields is high and the atomic system consists of several energy levels, this procedure can be quite cumbersome. A simpler way to drive the field equations is given by a Hamiltonian approach [48]. According to it, the polarization of the medium can be expressed as the partial derivative of the averaged free energy density of the atomic medium with respect to the electric field amplitude. In other words

$$P = -\left\langle \frac{\partial \hat{H}'}{\partial E^*} \right\rangle, \tag{50}$$

where $\hat{H}'$ is the interaction part of the Hamiltonian, $E^*$ is the complex amplitude of the electromagnetic field, and $P$ is the polarization of the medium. The fields that couple an initially prepared, collective, atomic state to other states of the atomic model are assumed to be very weak. Thus, the probability that, after the interaction, a state different from the initial one is populated is very small. This qualifies the initial state as a stationary state and the system will evolve in an adiabatic fashion, following its dynamics. Under these conditions, the averaged Hamiltonian that appears in Eq. (50) can be replaced by $E_{jm} - E^0{}_{jm}$. On the other hand, the total polarization of the BEC coupled to the probe electromagnetic field is given by

$$P = \varepsilon_0 \chi^{(1)} |E(\omega_p)| + \varepsilon_0 \chi^{(3)} |E(\omega_p)|^3 + \varepsilon_0 \chi^{(5)} |E(\omega_p)|^5 + ..., \tag{51}$$

where $\chi^{(1)}$ is the linear susceptibility, $\chi^{(k)}$ represents the $k$ th- order nonlinear susceptibility and $E(\omega_p)$ is the Fourier component of the mean value of the probe field at frequency $\omega_p$ defined as $E(\omega_p) = \varepsilon \sqrt{n}$, where $\varepsilon = \sqrt{\frac{\hbar \omega_p}{2\varepsilon_0 V}}$ ($V$ is the quantization volume). Hence we can rewrite Eq. (51) in the following form

$$P = \varepsilon_0 \chi^{(1)}(\omega_p)(\varepsilon\sqrt{n}) + \varepsilon_0 \chi^{(3)}(\omega_p)(\varepsilon\sqrt{n})^3 + \varepsilon_0 \chi^{(5)}(\omega_p)(\varepsilon\sqrt{n})^5 + .... \tag{52}$$

Using Eqs. (44) and (49) we find that

$$P = -\left\langle \frac{\partial H'}{\partial E^*} \right\rangle = -\left\langle \frac{\partial (E_{jm} - E^0{}_{jm})}{\partial \varepsilon \sqrt{n}} \right\rangle = \tag{53}$$

$$\varepsilon_0 \chi^{(1)}(\omega_p)(\varepsilon\sqrt{n}) + \varepsilon_0 \chi^{(3)}(\omega_p)(\varepsilon\sqrt{n})^3 + \varepsilon_0 \chi^{(5)}(\omega_p)(\varepsilon\sqrt{n})^5 + ....$$

From Eqs. (49) and (50) we obtain the following expression for the polarization of the BEC

$$P = -\left\langle \frac{\partial H'}{\partial E^*} \right\rangle = -\left\langle \frac{\partial (E_{jm} - E^0_{jm})}{\partial \varepsilon \sqrt{n}} \right\rangle =$$

$$= -\hbar\{(\frac{\omega_p}{2}+\Delta)+(\frac{3\omega_p}{2}+\Delta)(-\frac{1}{N}+\kappa)[-\frac{1}{2}+2(j+m)]-$$

$$K_1(\frac{\kappa}{2}-\frac{1}{2N})+K_2+K_2(\frac{\kappa}{2}-\frac{1}{2N})[\frac{1}{2}(j-m)(j-m-1)+\frac{1}{2}(j-m)]\}\sqrt{n} +$$  (54)

$$-\hbar\{(\frac{\omega_p+1}{2}+\Delta)(-\frac{1}{N}+\kappa)+2K_1(\frac{\kappa}{2}-\frac{1}{2N})+2K_2+K_2(\frac{\kappa}{2}-\frac{1}{2N})[-1-(j-m)]\}(\sqrt{n})^3 +$$

$$-\hbar\{\frac{3}{2}K_2(\frac{\kappa}{2}-\frac{1}{2N})\}(\sqrt{n})^5.$$

Using Eqs. (53) and (54) we obtain first, third and fifth order nonlinear susceptibilities of the BEC in the following form, and higher order susceptibilities of the medium are zero.

$$\chi^{(1)}(\omega_p) = \frac{-\hbar}{\varepsilon^2 \varepsilon_0}\{(\frac{\omega_p}{2}+\Delta)+(\frac{3\omega_p}{2}+\Delta)(-\frac{1}{N}+\kappa)[-\frac{1}{2}+2(j+m)]-$$

$$K_1(\frac{\kappa}{2}-\frac{1}{2N})+K_2+K_2(\frac{\kappa}{2}-\frac{1}{2N})[\frac{1}{2}(j-m)(j-m-1)+\frac{1}{2}(j-m)]\},$$  (55)

$$\chi^{(3)}(\omega_p) = \frac{-\hbar}{\varepsilon^4 \varepsilon_0}\{(\frac{3\omega_p}{2}+\Delta)(-\frac{1}{N}+\kappa)+2K_1(\frac{\kappa}{2}-\frac{1}{2N})+2K_2+K_2(\frac{\kappa}{2}-\frac{1}{2N})[-1-(j-m)]\},$$

$$\chi^{(5)}(\omega_p) = \frac{-\hbar}{\varepsilon^6 \varepsilon_0}\{\frac{3}{2}K_2(\frac{\kappa}{2}-\frac{1}{2N})\}.$$

The total susceptibility of the Bose condensate at the probe field frequency including linear and nonlinear terms, reads

$$\chi = \chi^{(1)}(\omega_p) + \chi^{(3)}(\omega_p)|E(\omega_p)|^2 + \chi^{(5)}(\omega_p)|E(\omega_p)|^4.$$  (56)

where the nonlinear part of the total susceptibility of the medium is

$$\chi^{(nl)}(\omega_p) = \chi^{(3)}(\omega_p)|E(\omega_p)|^2 + \chi^{(5)}(\omega_p)|E(\omega_p)|^4.$$
(57)

The refractive index $n(\omega_p)$ is related to the susceptibility of the medium $\chi(\omega_p)$ through the relation

$$n(\omega_p) = \sqrt{1+\chi(\omega_p)},$$  (58)

and the group refraction index is defined as

$$n_g = n(\omega_p) + \omega_p \frac{dn(\omega_p)}{d\omega_p}.$$
(59)

In the following we present the numerical results [Figs. 2-7] for the dependence of the refraction group index and the real and imaginary parts of the linear and nonlinear susceptibilities of the f- deformed BEC under consideration on the deformation parameters $\kappa$ and $N$ (or $\eta = \frac{1}{N}$). To show that our model leads to an efficient control of the group velocity of the probe pulse from subluminal to superluminal, we consider the BEC of sodium atoms as an example. Two hyperfine sub- levels of sodium state $3^2S_{1/2}$ with F=1 and F=2 are associated with levels $|2\rangle$ and $|1\rangle$ of the $\Lambda$- scheme, respectively. An excited state $|3\rangle$ corresponds to the hyperfine sub- level of the term $3^2P_{3/2}$ with F=2 (Fig. 1). The energy splitting between the levels $|1\rangle$ and $|2\rangle$ is denoted by $\frac{\omega_{12}}{2\pi} = 1772 MHz$ and the transition $|3\rangle \to |2\rangle$ corresponds to the optical frequency $\frac{\omega}{2\pi} = 5.1 \times 10^{14} Hz$ [7]. We choose the density of the condensate $\frac{N}{V} = 3.3 \times 10^{12} cm^{-3}$ [7], the dipole matrix element $|\mu_{32}| = 22 \times 10^{-30} C.m$ [49], the decay rates $\gamma_{31}$ and $\gamma_{32}$ of the level $|3\rangle$ $\gamma_{31}/2\pi = \gamma_{32}/2\pi = 5 MHz$ [48], the decay rates from the transition between the hyperfine levels $|1\rangle$ and $|2\rangle$ $\gamma_{12}/2\pi = 38 KHz$ [49], the coupling constant $g_1/2\pi = 21.4 MHz$ [7] and the intensity of the probe pulse $I_p = \frac{80 \mu W}{cm^2}$ which corresponds to 25 photons on average and the coupling intensity $I_c = \frac{55 mW}{cm^2}$.

In Fig. 2a, we plot the group refraction index as a function of the detuning parameter $\Delta$ for three different values of the deformation parameter $\kappa$. We set $N = 10^{14}, (\eta \approx 0)$ and investigate the effects of the collisions between the atoms within the condensate. The point $\Delta = 0$ of the exact resonance corresponds to the EIT regime, characterized by low losses. The zero points of group refraction index in Fig. 2a, show an uncertainty of the group velocity $v_g$, of the probe pulse and describes the possibility of observing superluminal velocities. When the refraction index is negative (corresponding to the superluminal propagation) the peak of the pulse exits the BEC before it passes the entrance face. However, with increasing the collision rate $\kappa$ the subluminal and superluminal group velocity is enhanced. The physical origin of this result is that with increasing the deformation parameter $\kappa$ the nonlinearity of the model under consideration becomes larger, as shown in Fig. 2b. In Fig. 2b, we plot the real part of the nonlinear total susceptibility $\chi^{(nl)}$ [Eq. 57] as a function of $\Delta$ for three different values of $\kappa$. The real part of the complex $\chi^{(nl)}$ shows the nonlinearity of this process. It is evident that the nonlinearity increases with deformation parameter $\kappa$. To observe the effect of number of atoms $N$, for a given collision rate, we plot the group refraction index as a function of detuning $\Delta$ for different values of $N$ (Fig. 3a). We see that the pulse propagation can change from subluminal to superluminal and significant superluminality and subluminality takes place for large values of the deformation parameter $\eta$. In other words, superluminality and subluminality decrease with the total number of condensate atoms $N$. The reason is due to the fact that with increasing the deformation parameter $\eta$ there exist an enhancement of the nonlinearity as shown in Fig. 3b. In Fig. 3b, we plot the real part of the total nonlinear susceptibility $\chi^{(nl)}$ as a function of detuning $\Delta$ for different values of the deformation parameter $\eta$ and for the case of no collision ($\kappa = 0$). As is seen, the nonlinearity increases with deformation parameter $\eta$. The imaginary part of the total nonlinear susceptibility $\chi^{(nl)}$ is plotted versus the detuning $\Delta$ in Figs. 4a and 4b. We observe that the absorption coefficients increase with the deformation parameters $\eta$ and $\kappa$. In Figs. 5a and 5b, we plot the real part of the linear susceptibility $\chi^{(1)}$ versus detuning $\Delta$ for various values of the parameters $\eta$ and $\kappa$. It is evident that the dispersive properties of the BEC increase with the deformation parameter $\eta$ and $\kappa$. The imaginary part of the linear susceptibility as a function of the detuning $\Delta$ is shown in Figs. 6a and 6b, for various values of the parameter $\eta$ and $\kappa$. It is clear that the medium has absorption for positive absorption coefficient and gain for negative absorption coefficient. We see that the gain and absorption properties of the BEC increase with the deformation parameters $\eta$ and $\kappa$. In figs. 7a-7d the real and imaginary parts of the total susceptibility [Eq. (56)] are plotted versus the detuning parameter $\Delta$, which correspond to the dispersive and absorptive properties of the condensate, respectively. It is seen that at the region around the zero detuning $\Delta$ both $\chi'$, $\chi''$ are equal to zero. This means that the absorption is almost zero where the index of refraction is unity. Thus the medium becomes transparent under the action of probe field and EIT is occurred. It is evident that with increase of the parameters $\eta$ and $\kappa$ there exists an enhancement of the dispersion and absorption properties of the deformed BEC under consideration.

## 5 Summary and conclusions

In summary, we have studied the propagation of a weak optical probe field in an f-deformed BEC of the gas of $\Lambda$- type three- level atoms in the EIT regime. We have employed the quantum theory of the angular momentum to obtain the eigenvalues and eigenfunctions of the model up to first- order approximation. By applying an effective two- level quantum model within the framework of the f- deformed boson model, we have found an explicit expression for the linear and nonlinear susceptibilities of the atomic condensates. We have seen that the particle- number conservation in BEC requires a deformation on the bosonic field. In addition, we have considered the effects of collisions between the atoms within condensate as a special kind of f- deformation for which the collision rate $\kappa$ is regarded as the corresponding deformation parameter. We have demonstrated that tunable control of the group velocity of a weak probe field from subluminal to superluminal. We have found that the deformed parameters $\eta$ and $\kappa$ play an important role in determining the subluminal and superluminal propagation through the condensate. We have also shown that by applying the deformation on atomic operators of BEC medium, it is possible to obtain large nonlinearity that leads to an enhanced subluminal and superluminal propagation.

**Figure captions**

Fig.1: The energy level $\Lambda$ scheme of three-level atoms

Fig.2a: The group velocity as a function of detuning $\Delta$, for different values of the deformation parameter $\kappa$, $\kappa = 0(...), \kappa = 0.005\text{Hz}(-), \kappa = 0.008\text{Hz}(-.-.)$ and for the total number of atoms $N = 10^{14}, (\eta \approx 0)$ and the total number of excited atoms $n_e = 1$

Fig.2b: The real part of the total nonlinear susceptibility $\text{Re}(\chi^{(nl)})$, as a function of detuning $\Delta$, for three different values of the deformation parameter $\kappa$, $\kappa = 0(...), \kappa = 0.005\text{Hz}(-), \kappa = 0.008\text{Hz}(-.-.)$ and for the total number of atoms $N = 10^{14}, (\eta \approx 0)$ and the total number of excited atoms $n_e = 1$

Fig.3a: The group velocity as a function of detuning $\Delta$, for three different values of the total number of atoms $N$, $N = 300(...), N = 200(-), N = 100(-.-.)$ in the case of no collision ($\kappa = 0$), and the total number of excited atoms $n_e = 1$

Fig.3b: The real part of the total nonlinear susceptibility $\text{Re}(\chi^{(nl)})$, as a function of detuning $\Delta$, for three different values of the total number of atoms $N$, $N = 300(...), N = 200(-), N = 100(-.-.)$ in the case of no collision ($\kappa = 0$), and the total number of excited atoms $n_e = 1$

Fig.4a: The imaginary part of the total nonlinear susceptibility $\text{Im}(\chi^{(nl)})$, as a function of detuning $\Delta$, for three different values of the deformation parameter $\kappa$, $\kappa = 0(...), \kappa = 0.005\text{Hz}(-), \kappa = 0.008\text{Hz}(-.-.)$ and for the total number of atoms $N = 10^{14}, (\eta \approx 0)$ and the total number of excited atoms $n_e = 1$

Fig.4b: The imaginary part of the total nonlinear susceptibility $\text{Im}(\chi^{(nl)})$, as a function of detuning $\Delta$, for three different values of the total number of atoms $N$, $N = 300(...), N = 200(-), N = 100(-.-.)$ in the case of no collision ($\kappa = 0$), and the total number of excited atoms $n_e = 1$

Fig.5a: The real part of the linear susceptibility $\text{Re}(\chi^{(1)})$ as a function of detuning $\Delta$, for three different values of the deformation parameter $\kappa$, $\kappa = 0(...), \kappa = 0.005\text{Hz}(-), \kappa = 0.008\text{Hz}(-.-.)$ and for the total number of atoms $N = 10^{14}, (\eta \approx 0)$ and the total number of excited atoms $n_e = 1$

Fig.5b: The real part of the linear susceptibility $\operatorname{Re}(\chi^{(1)})$ as a function of detuning $\Delta$, for three different values of the total number of atoms $N$, $N=300(...), N=200(-), N=100(-.-.)$ in the case of no collision ($\kappa=0$), and the total number of excited atoms $n_e=1$

Fig.6a: The imaginary part of the linear susceptibility $\operatorname{Im}(\chi^{(1)})$ as a function of detuning $\Delta$, for three different values of the deformation parameter $\kappa$, $\kappa=0(...), \kappa=0.005\text{Hz}(-), \kappa=0.008\text{Hz}(-.-.)$ and for the total number of atoms $N=10^{14}, (\eta \approx 0)$ and the total number of excited atoms $n_e=1$.

Fig.6b: The imaginary part of the linear susceptibility $\operatorname{Im}(\chi^{(1)})$ as a function of detuning $\Delta$, for three different values of the total number of atoms $N$, $N=300(...), N=200(-), N=100(-.-.)$ in the case of no collision ($\kappa=0$), and the total number of excited atoms $n_e=1$

Fig.7a: The real part of the susceptibility $\operatorname{Re}(\chi)$, as a function of detuning $\Delta$, for three different values of the deformation parameter $\kappa$, $\kappa=0(...), \kappa=0.005\text{Hz}(-), \kappa=0.008\text{Hz}(-.-.)$ and for the total number of atoms $N=10^{14}, (\eta \approx 0)$ and the total number of excited atoms $n_e=1$

Fig.7b: The imaginary part of the susceptibility $\operatorname{Im}(\chi)$, as a function of detuning $\Delta$, for three different values of the deformation parameter $\kappa$, $\kappa=0(...), \kappa=0.005\text{Hz}(-), \kappa=0.008\text{Hz}(-.-.)$ and for the total number of atoms $N=10^{14}, (\eta \approx 0)$ and the total number of excited atoms $n_e=1$

Fig.7c: The real part of the susceptibility $\operatorname{Re}(\chi)$, as a function of detuning $\Delta$, for three different values of the total number of atoms $N$, $N=300(...), N=200(-), N=100(-.-.)$ in the case of no collision ($\kappa=0$), and the total number of excited atoms $n_e=1$

Fig.7d: The imaginary part of the susceptibility $\operatorname{Im}(\chi)$, as a function of detuning $\Delta$, for three different values of the total number of atoms $N$, $N=300(...), N=200(-), N=100(-.-.)$ in the case of no collision ($\kappa=0$), and the total number of excited atoms $n_e=1$

Fig. 1

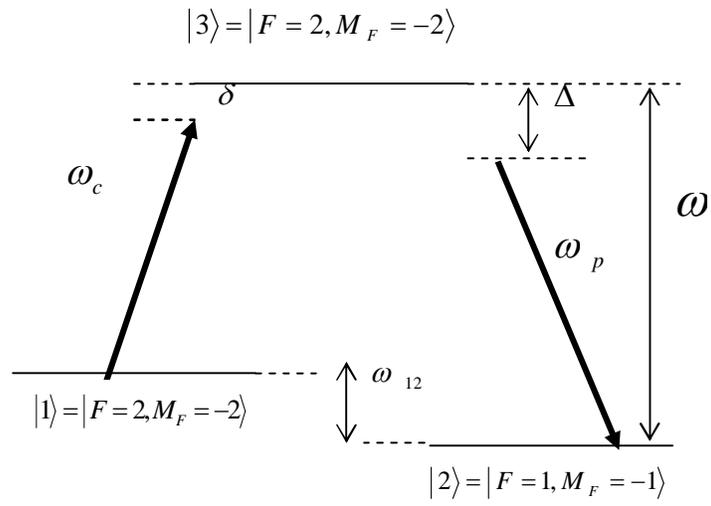

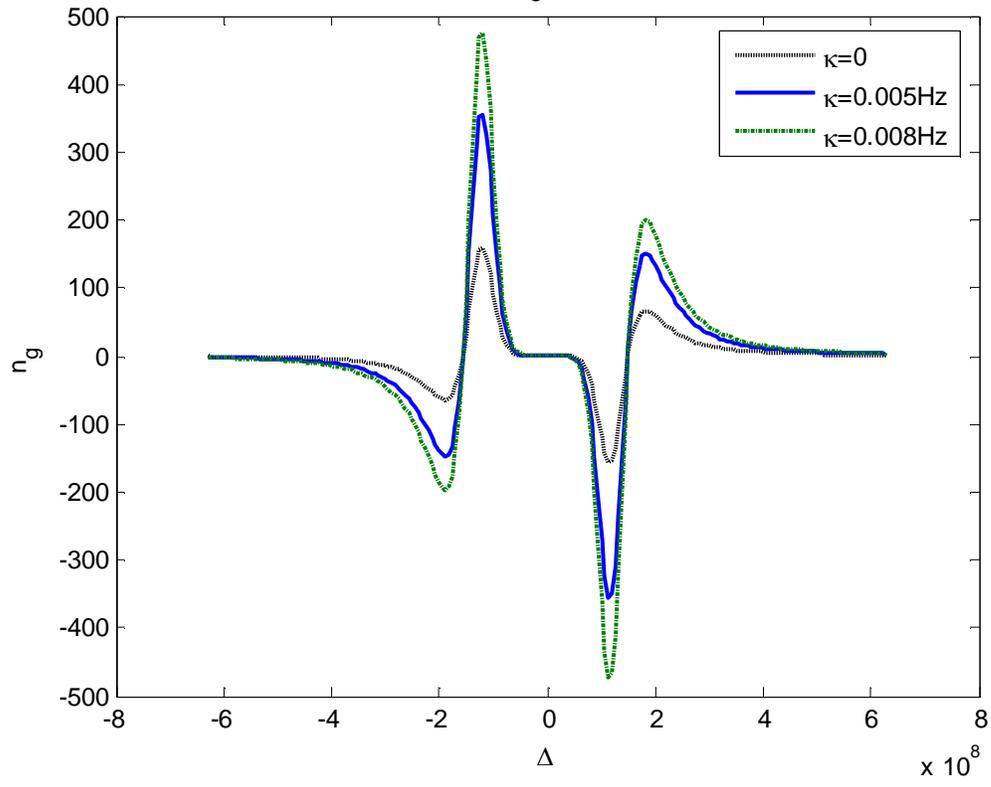

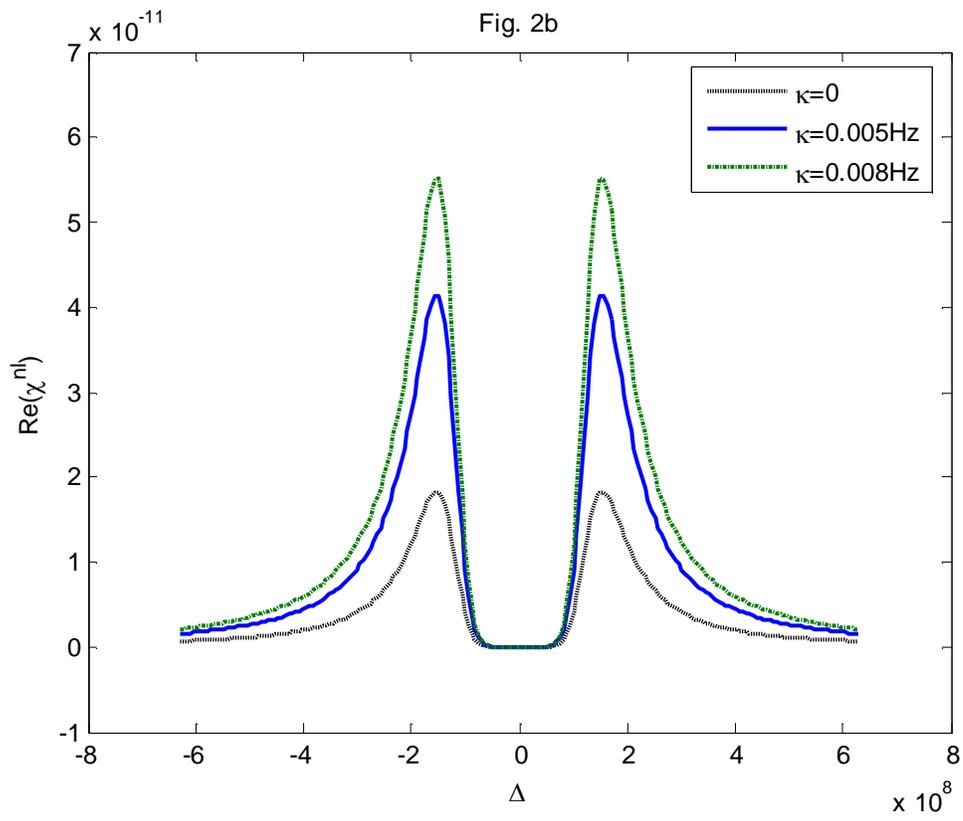

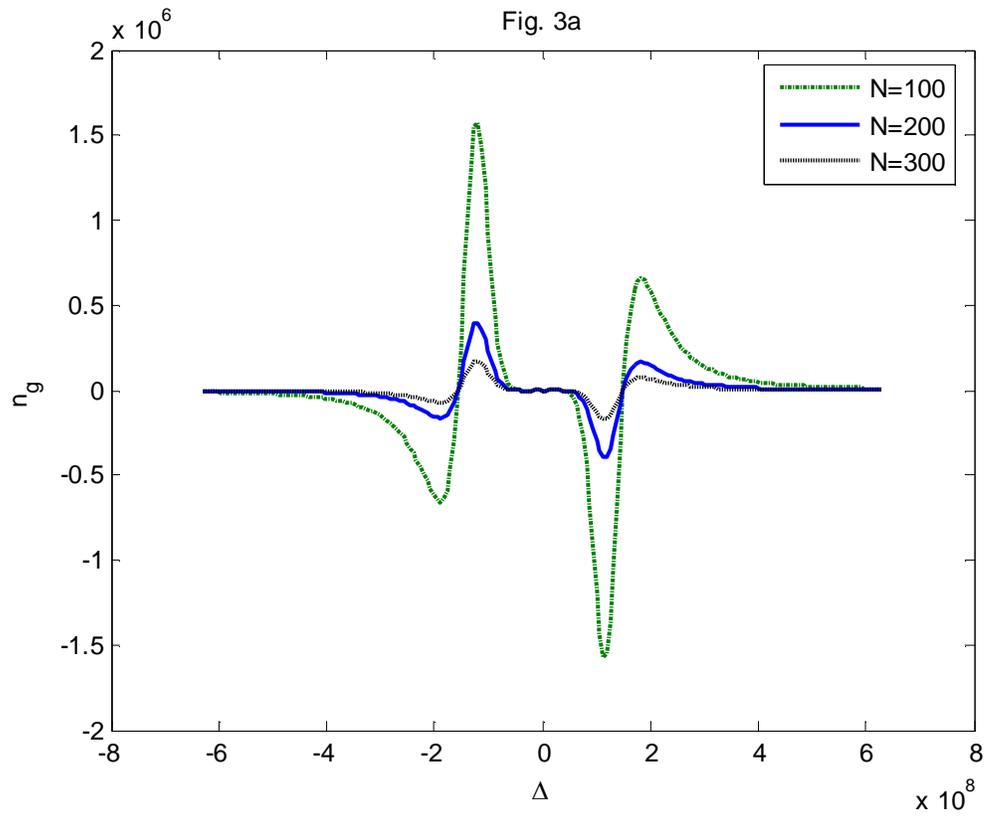

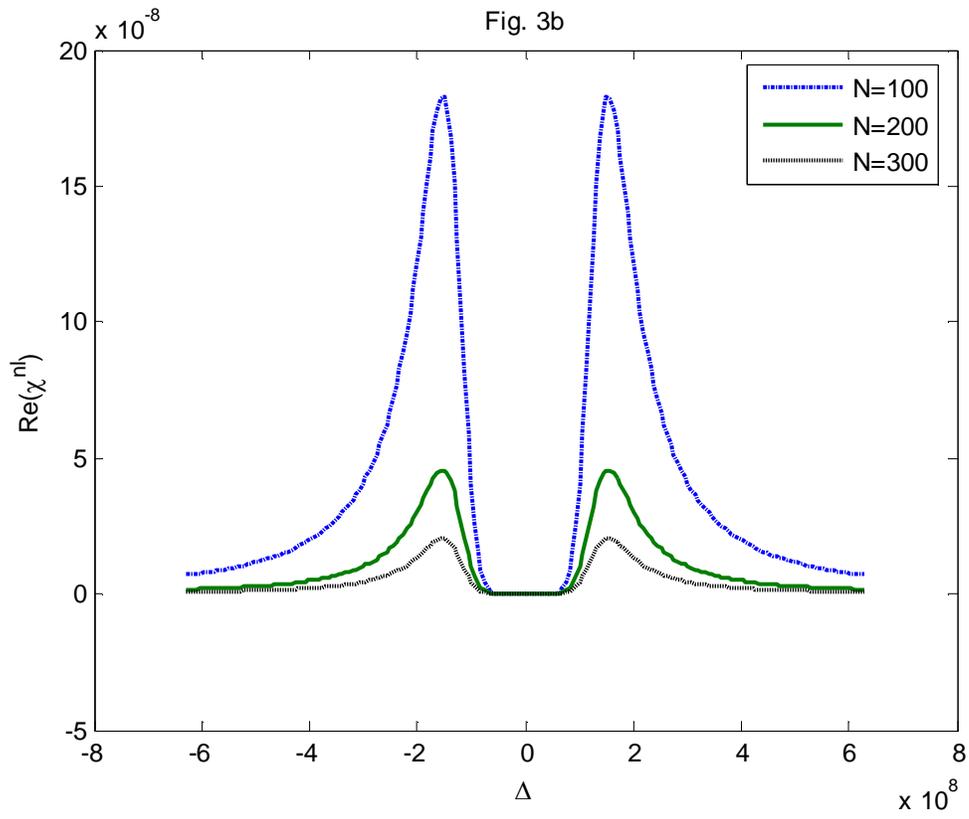

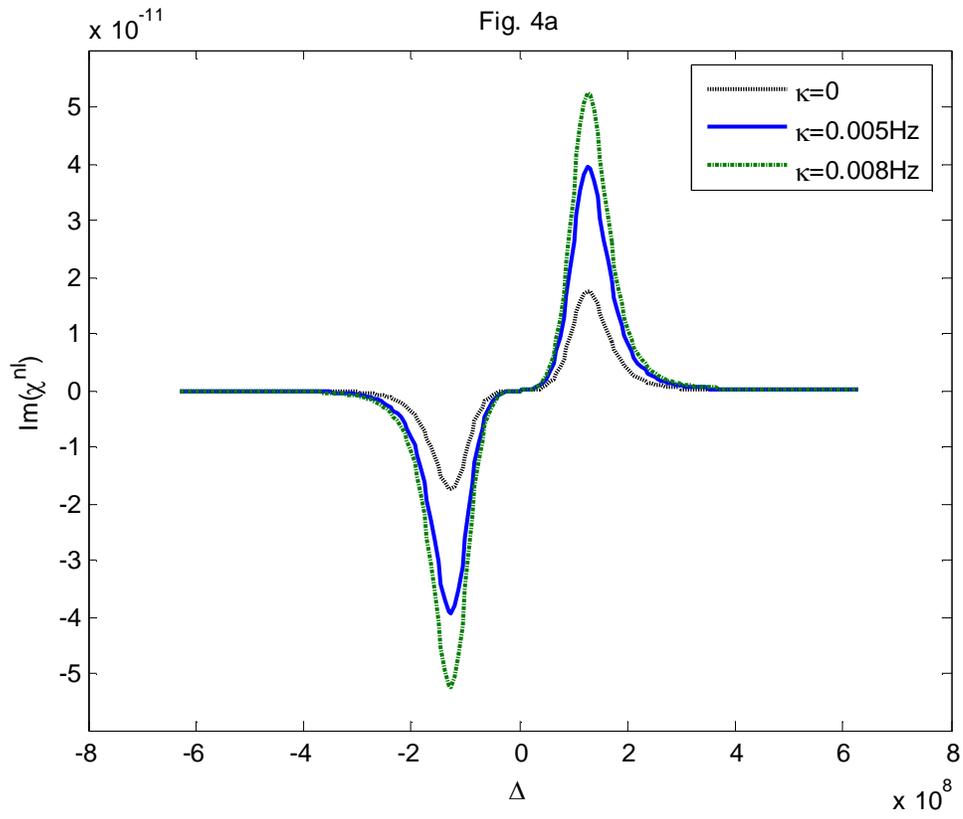

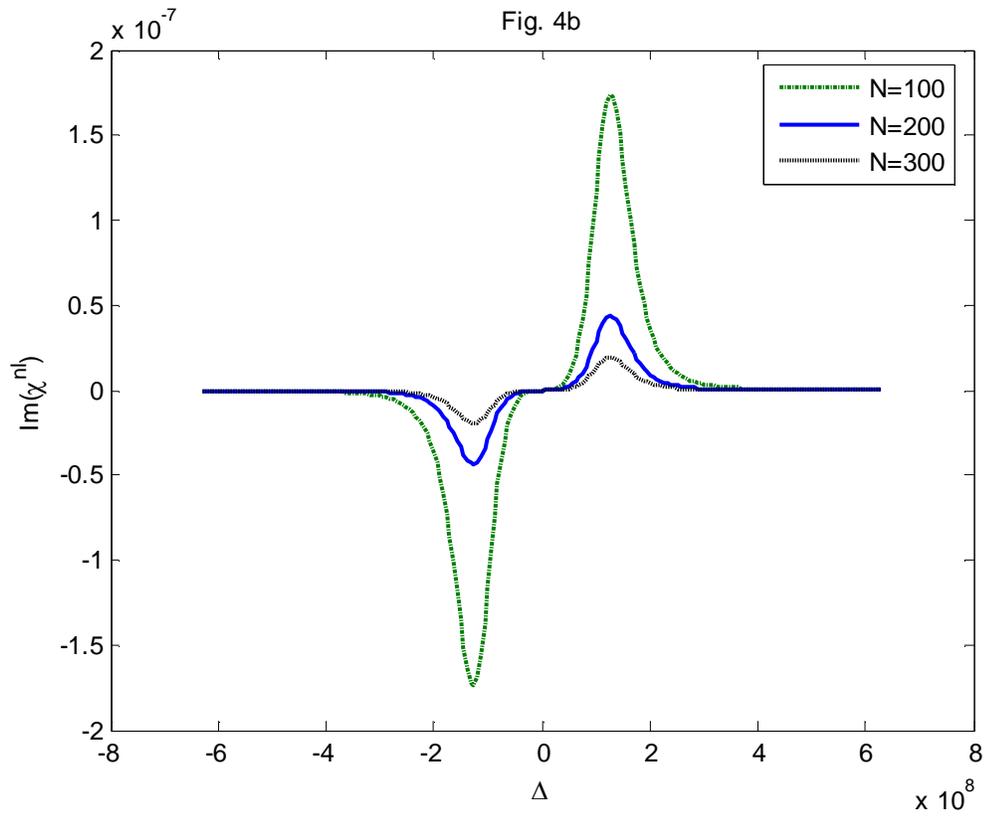

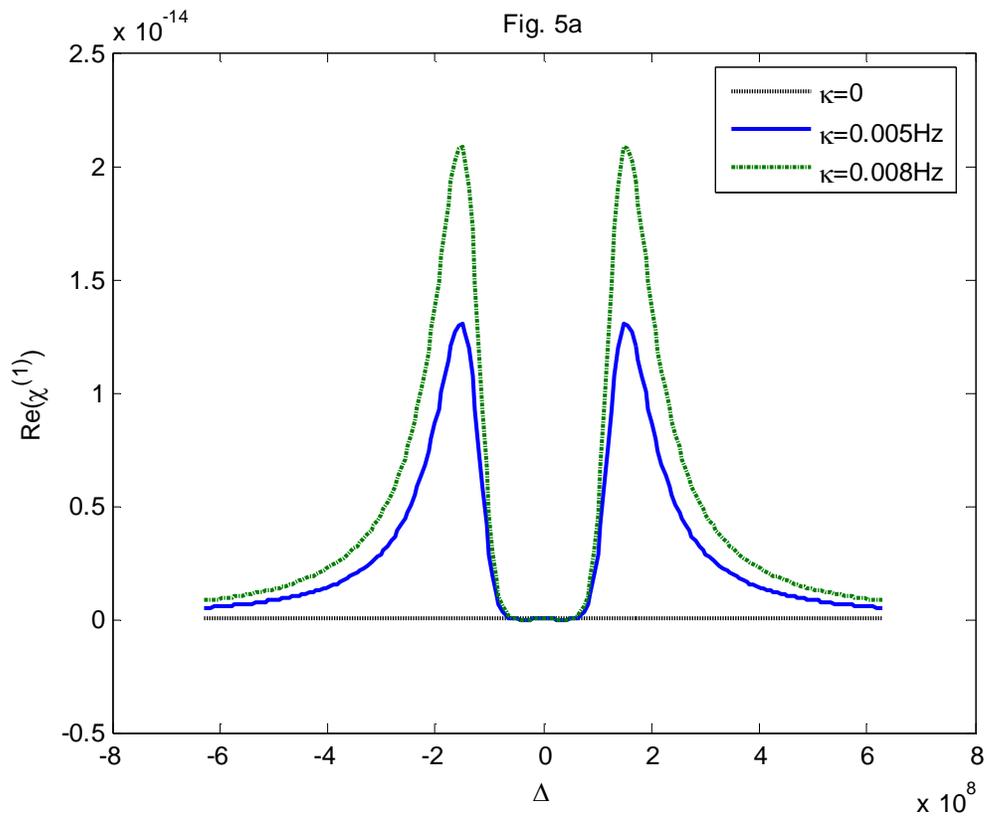

Fig. 5a

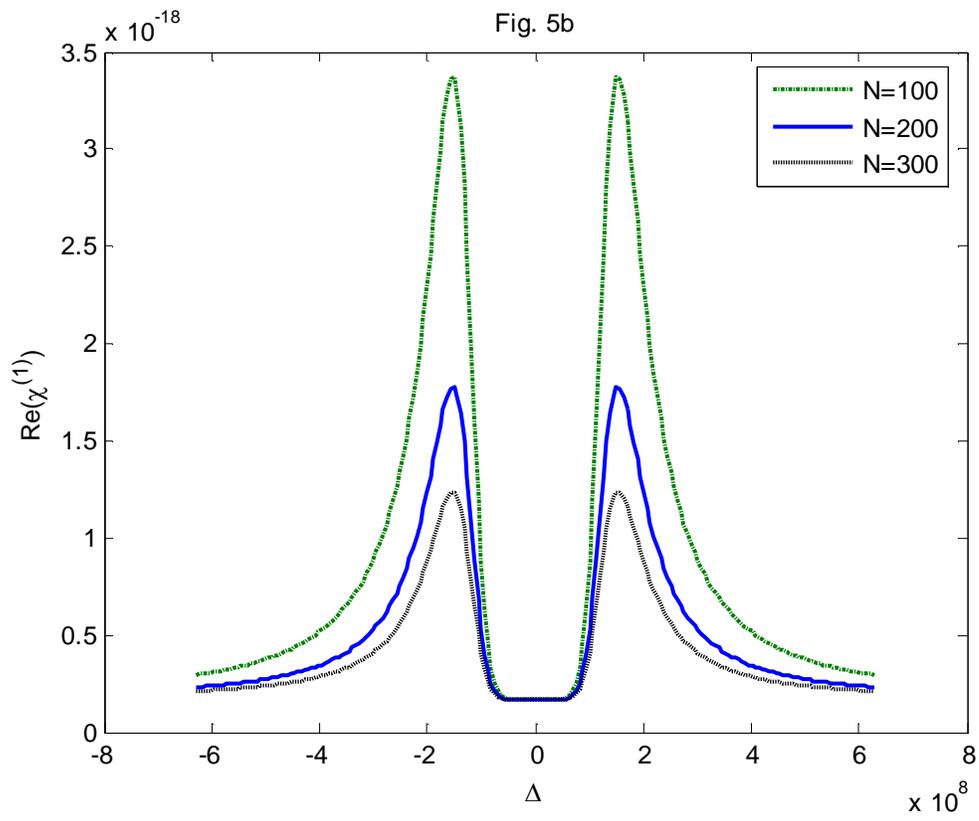

Fig. 5b

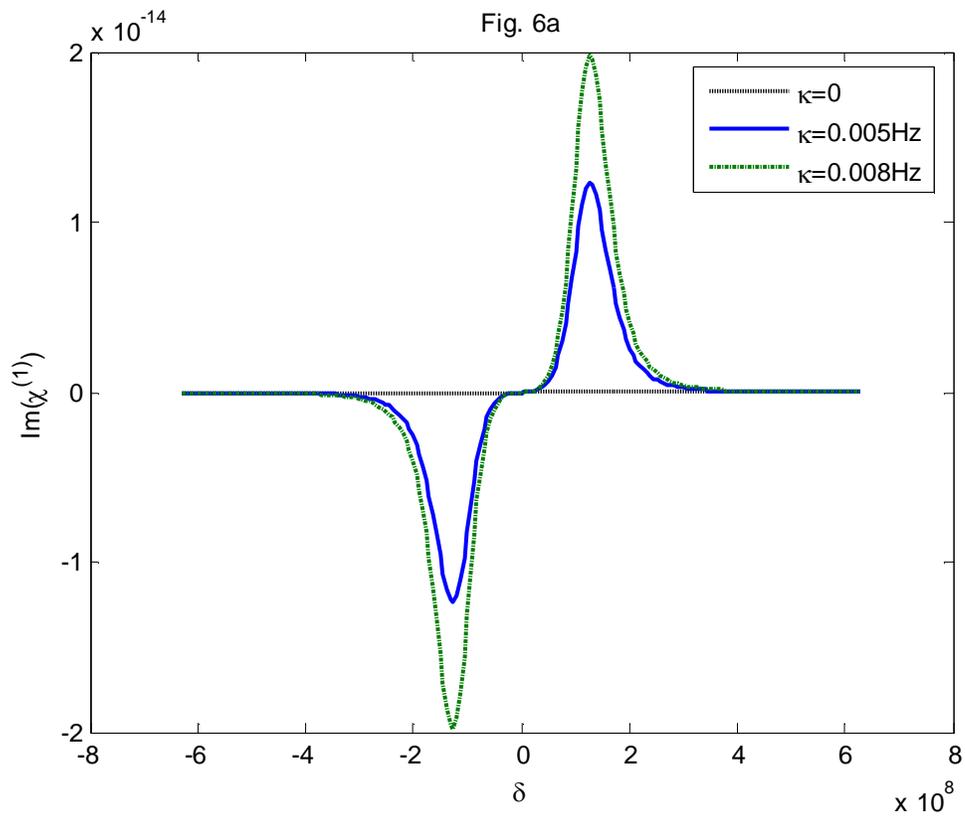

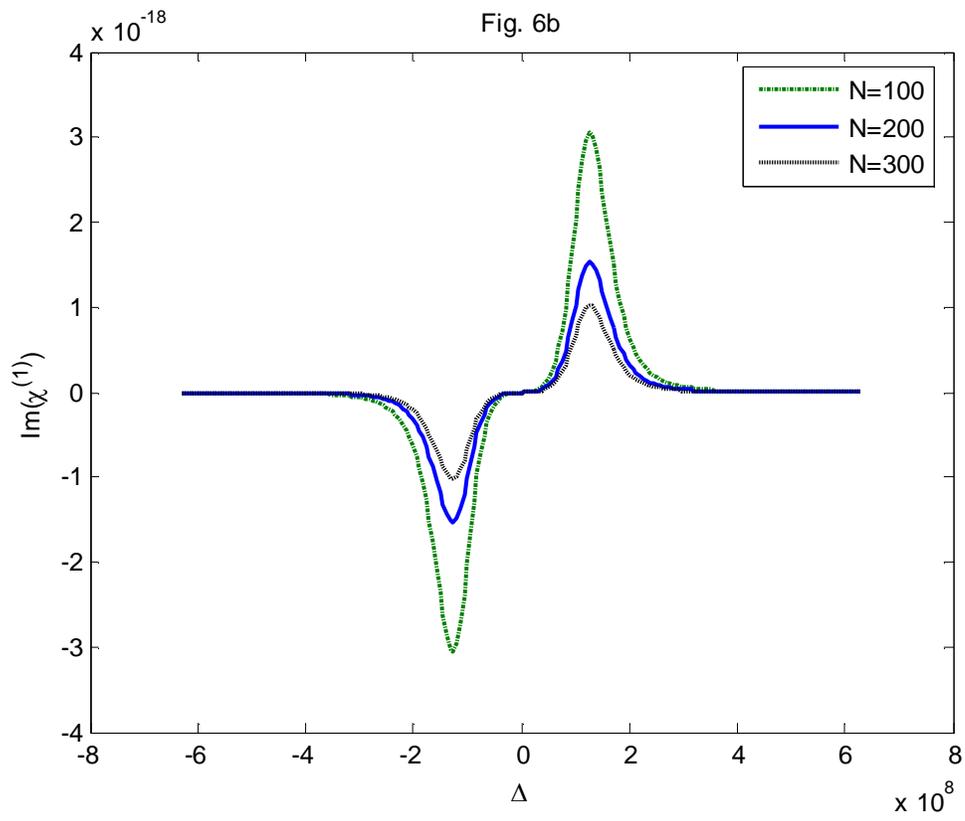

Fig. 6b

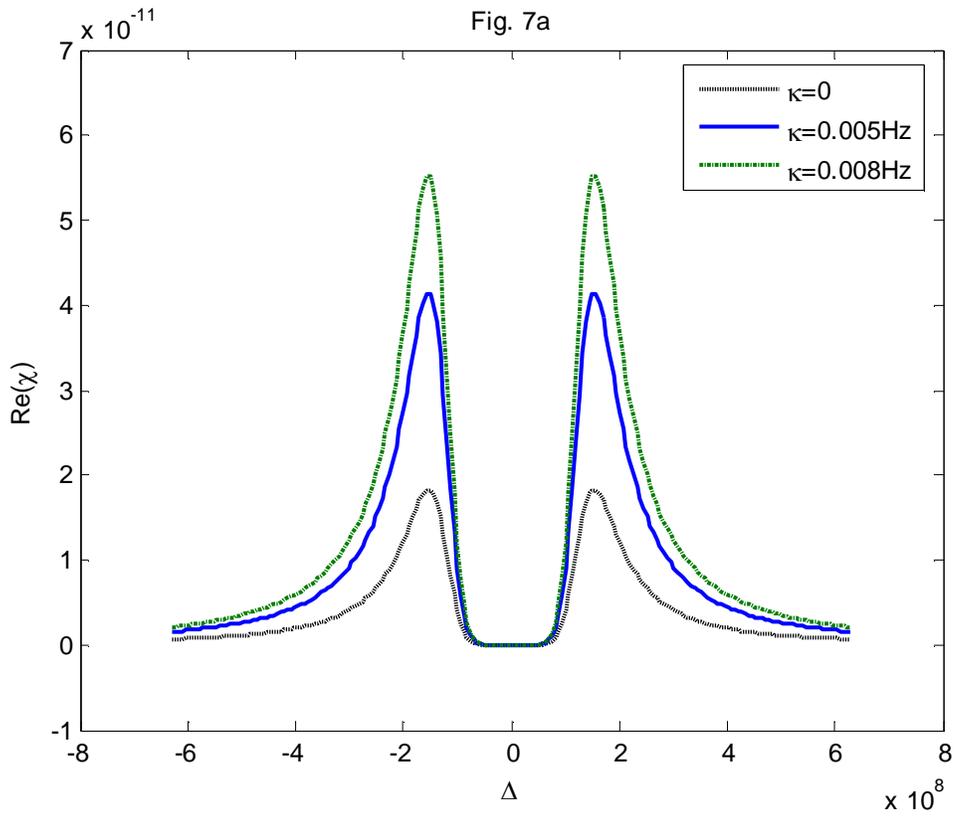

Fig. 7a

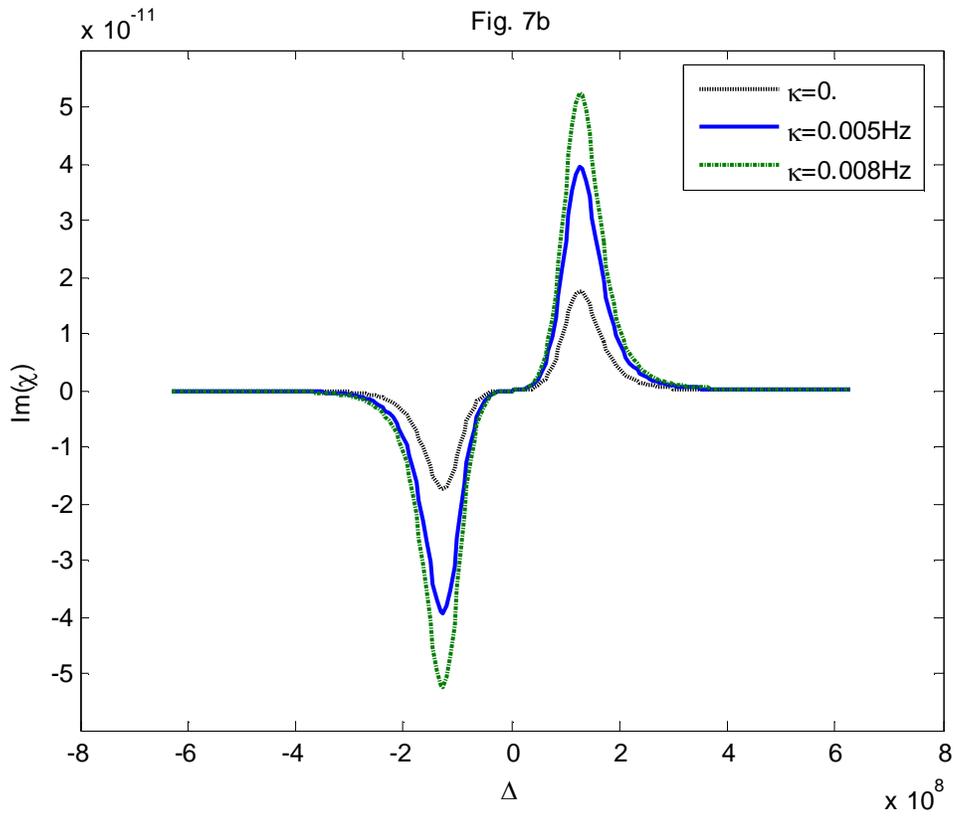

Fig. 7b

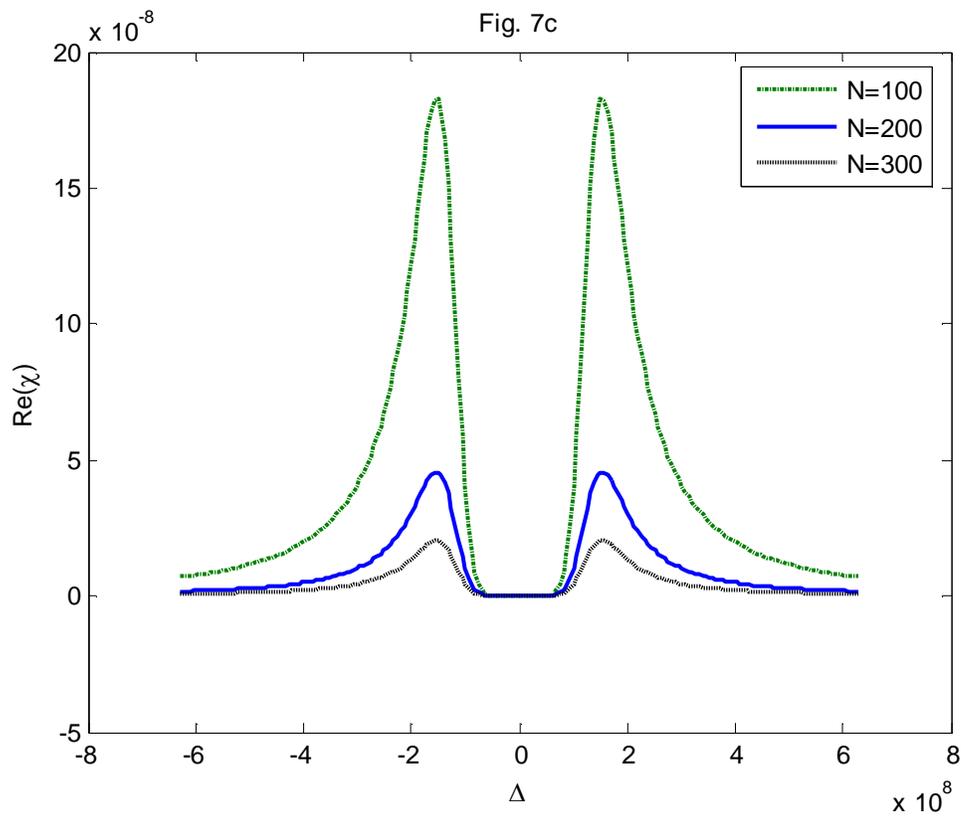

Fig. 7c

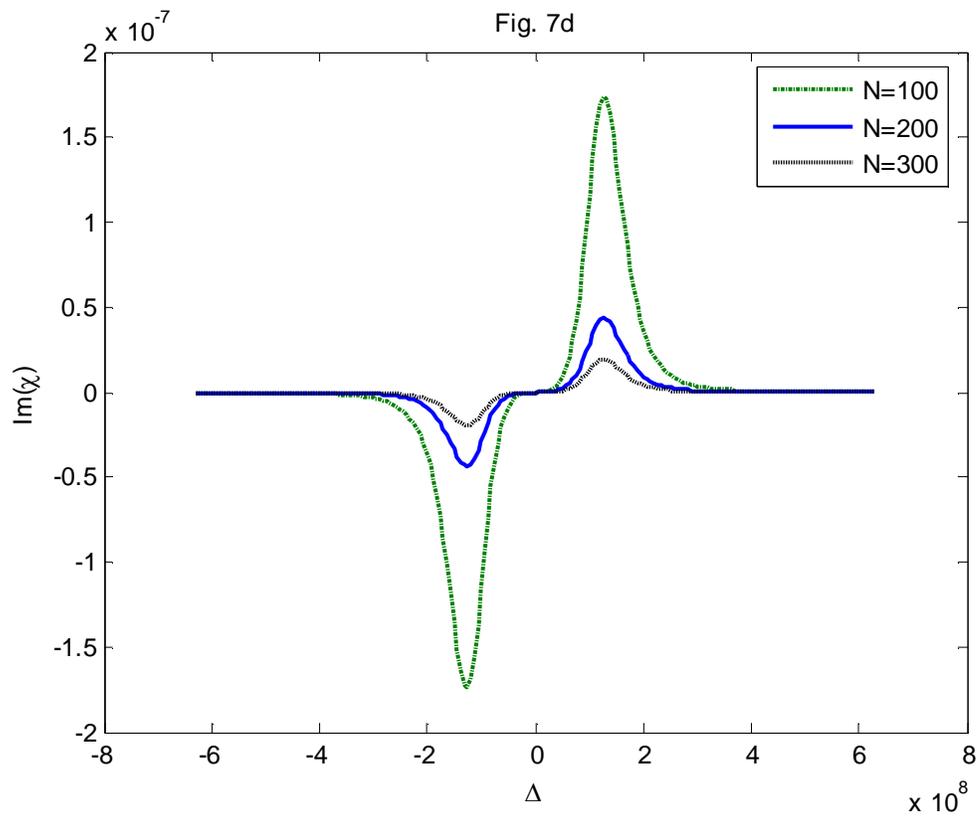
Fig. 7d